\PassOptionsToPackage{dvipsnames}{xcolor}
\documentclass[manuscript, screen, review=false, pbalance, anonymous=false, nonacm]{acmart}
\usepackage{soul}
\usepackage{tablefootnote}
\usepackage{svg}

\usepackage{booktabs}
\usepackage{todonotes}
\usepackage{listings}
\definecolor{LightGray}{rgb}{0.97,0.97,0.97}
\lstdefinelanguage{SPARQL}{
  basicstyle=\small\ttfamily,
  backgroundcolor=\color{LightGray},
  columns=fullflexible,
  breaklines=false,
  sensitive=true,
  frame=bt,
  aboveskip=1em,
  belowskip=1em,
  xleftmargin=.5em,
  xrightmargin=.5em,
  framexleftmargin=.5em,
  framextopmargin=.5em,
  framexbottommargin=.5em,
  framexrightmargin=.5em,
  tabsize = 2,
  showstringspaces=false,
  morecomment=[n][\color{blue}]{<http}{>}, 
  morestring=[b][\color{OliveGreen}]{\"},  
  keywordsprefix=?,
  classoffset=0,
  keywordstyle=\color{Sepia},
  morekeywords={},
  classoffset=1,
  keywordstyle=\color{Purple},
  morekeywords={rdf,rdfs,owl,xsd,purl,lsc},
  classoffset=2,
  keywordstyle=\color{MidnightBlue},
  morekeywords={
    SELECT,CONSTRUCT,DESCRIBE,ASK,WHERE,FROM,NAMED,PREFIX,BASE,OPTIONAL,
    FILTER,GRAPH,LIMIT,OFFSET,SERVICE,UNION,EXISTS,NOT,BINDINGS,MINUS,a,
    BIND,GROUP,ORDER,BY,CONTAINS,DESC,COUNT,AS
  }
}

\usepackage[capitalise,noabbrev]{cleveref}
\usepackage{graphicx}
\usepackage{csquotes}
\usepackage[htt]{hyphenat}
\usepackage{multirow}
\usepackage{tabularx}
\usepackage[normalem]{ulem}
\usepackage{mathtools}

\newcommand{\quot}[1]{“#1”}

\newcommand{\megras}[0]{\emph{MeGraS}}
\newcommand{\ligr}[1]{\anon[Anonymous Application (Version #1)]{\emph{LifeGraph~#1}}}
\newcommand{\ligra}[0]{\emph{LifeGraph}}
\newcommand{\ums}[0]{\emph{Unified Multimedia Segmentation}}
\newcommand{\mg}[0]{\emph{MediaGraph}}

\begin{document}

\title[MediaGraph: A Content-Aware Data Model and Query Framework for Multimodal Knowledge Graphs]{MediaGraph: A Content-Aware Data Model and Query Framework for Multimodal Knowledge Graphs}

\author{Florian Ruosch}
\orcid{0000-0002-0257-3318}
\affiliation{%
  \institution{FPR Consulting}
  \city{Zurich}
  \country{Switzerland}
}
\email{fpr_consulting@hispeed.ch}

\author{Luca Rossetto}
\orcid{0000-0002-5389-9465}
\affiliation{%
  \department{School of Computing}
  \institution{Dublin City University}
  \city{Dublin}
  \country{Ireland}
}
\email{luca.rossetto@dcu.ie}

\begin{abstract}
    Multimodal knowledge graphs typically treat multimedia documents as opaque, external entities.
    This \emph{content-agnostic} approach constrains retrieval and analysis by isolating media from the graph's core structure, hindering the ability to capture and query complex relationships across media types.
    To address this, we introduce the \textbf{\textit{MediaGraph} Data Model} and its prototypical implementation \textbf{\megras}, the \textbf{Me}dia\textbf{Gra}ph \textbf{S}tore, a novel approach that integrates multimedia content as graph nodes.
    This paradigm shift enables the query engine to directly access and process a document's intrinsic content, allowing for native operations such as feature-based similarity search, dynamic segmentation, and the inference of non-materialized relations.
    By extending the SPARQL query language, \megras{} provides a cohesive platform for the storage, management, and expressive querying of multimodal data.
    \megras{} is open-source software that establishes a new framework, moving the field toward \emph{content-aware} multimodal knowledge graphs.
\end{abstract}



\maketitle


\section{Introduction}
\label{sec:IN}

The effective modeling and querying of the world's ever-growing collection of digital multimedia is a significant challenge~\cite{giangreco2018database}.
Multimodal data, such as images, video, audio, and text, present a richness that traditional information management systems struggle to capture~\cite{chen2024knowledge}.
Knowledge graphs are a useful and widely adopted method to store structured information, but they often separate easily represented data (such as properties) from the multimedia documents themselves~\cite{DBLP:journals/tkde/ZhuLWJSWXY24}.
Multimodal knowledge graphs incorporate multimedia documents via external Uniform Resource Identifier (URI) references or by storing them as \quot{opaque blobs,} treating them in a fundamentally \quot{content-agnostic} way.
This approach severely limits query expressiveness because the underlying engine lacks direct access to node contents, hindering deeper analysis, making the computation of rich, cross-modal queries impossible, and obscuring complex real-world relationships between multimedia documents.
This is a major hurdle in advancing the fields of multimedia modeling and retrieval, as well as their applications, such as semantic memory retrieval~\cite{lsc25} or interactive video exploration~\cite{DBLP:journals/mms/LokocABDGMMNPRSSSKSVV23}.

The root cause of these limitations is not merely an implementation gap, but a \emph{conceptual} one: existing data models for multimodal knowledge graphs were not designed to give the graph and, more importantly, its query engine meaningful access to the intrinsic content of media.
To address this, we introduce the \textbf{\mg{} Data Model}, a formal framework that dissolves the barrier between structured symbolic knowledge and unstructured multimedia content.
Rather than treating media as atomic, unsearchable units, the \mg{} model elevates multimedia documents and their segments to \quot{first-class citizens} in the graph, natively enabling operations such as k-nearest neighbor search, dynamic segmentation, and the inference of non-materialized relations derived from content at query time.
We characterize this as a paradigm shift from \quot{content-agnostic} to \quot{content-aware}~\anon[(Anonymous Reference)]{\cite{ruosch2026applications}}, a move essential to advancing the field of multimedia modeling.
This new paradigm offers unprecedented granularity and flexibility for complex content analysis and retrieval in multimodal knowledge graphs.

To validate and instantiate this conceptual framework, we present \textbf{\megras}~\anon[(Anonymous Reference)]{\cite{Rossetto:2025:MGR}}, short for \textbf{Me}dia\textbf{Gra}ph \textbf{S}tore, an open-source store and query engine built on the \mg{} Data Model.
\megras{} provides a concrete, working realization of the content-aware paradigm, extending the SPARQL query language to support the model's full expressive capabilities.
It is released as open source to serve as a research platform, enabling the community to explore, build upon, and extend content-aware multimodal knowledge graphs.

The core contributions of this article are as follows:
\begin{itemize}
    \item \textbf{The \mg{} Data Model}:
        We introduce a novel data model specifically designed to capture the richness and complexity of multimodal data, integrating media directly into the knowledge graph's structure rather than treating them as opaque external BLOBs.
        The model enables access to the contents and internal structures of media as first-class, URI-addressable citizens in the graph, alongside semantic relationships.
        It provides a formal, extensible framework for representing not only entire documents but also their constituent segments, enabling more granular, expressive querying than previously possible.
        In doing so, it introduces and formalizes the \quot{content-aware} paradigm as a principled alternative to the content-agnostic status quo.

    \item \textbf{\ums{} and Feature Extraction}:
        The \mg{} model's content-aware perspective is operationalized through integration of the \ums~\cite{DBLP:journals/tgdk/WilliBR24}, which provides a formal foundation for decomposing diverse media types into addressable segments under a consistent URI naming scheme.
        We further describe how direct document access enables feature extraction, yielding additional relations such as extracted content (e.g., video frames or document spans) and properties derived from them (e.g., embeddings).
        These are the key processes by which a \mg{} is populated with the rich interconnections that enable complex querying and analysis.

    \item \textbf{Cross-Modal Analysis}:
        We demonstrate how the \mg{} model enables querying relationships that span disparate media types, using structured knowledge to connect seemingly unrelated media by inferring links not only between whole documents but also between their internal contents.
        For example, the same person can be identified in an image and a video and then linked to a recording of their voice; or text extracted from a document can be embedded to retrieve an image it describes.
        This enables a class of complex, cross-modal queries that is fundamentally inaccessible to content-agnostic systems.

    \item \textbf{Advanced Multimedia Retrieval and Inference}:
        We describe a content-aware query framework that extends SPARQL to enable a new class of multimedia queries.
        Instead of being limited to simple tag-based searches, users can formulate complex, expressive queries that directly relate to the properties and contents of multimedia documents.
        Central to this is the distinction between \emph{implicit relations}, such as k-nearest neighbor associations, which are evaluated at query time but never persisted, and \emph{derived relations}, such as embeddings or OCR transcripts, which are computed on demand and then stored for future use.

    \item \textbf{An Open-Source Framework}:
        We present \megras{} as an open-source instantiation of the \mg{} Data Model, publicly available as both a compilable research prototype and a Docker container to facilitate adoption and extension by the research community.\footnote{\label{fn:repo}\anon{\url{https://megras.org}}}\textsuperscript{,}\footnote{\label{fn:docker}\anon{\url{https://megras.org/docker}}}

    \item \textbf{Use Cases and Applications}:
        To illustrate the practical value of the content-aware paradigm, we describe use cases in which \megras{} has demonstrated its capabilities, including participation in the Lifelog Search Challenge~\cite{lsc25} and the modeling of scientific literature.
        We also outline a range of potential future applications.
\end{itemize}


\section{Related Work}
\label{sec:RW}

The foundations of \megras{} are built upon decades of research in knowledge representation and multimedia analysis.
This section reviews the current state of multimodal knowledge graphs and multimedia information retrieval, highlighting the technical challenges in scalability, query expressiveness, and the \quot{semantic gap.}

\subsection{Multimodal Knowledge Graphs}

Research in knowledge representation has recently shifted toward integrating heterogeneous data types, giving rise to multimodal knowledge graphs in two primary structural paradigms: attribute-based and node-based models~\cite{DBLP:journals/tkde/ZhuLWJSWXY24}.
In the attribute-based approach, multimodal data such as images, audio, or video are treated as literal attribute values or simple URIs attached to an entity, serving primarily as descriptive metadata.
In contrast, node-based multimodal knowledge graphs treat multimedia documents as central entities or nodes within the graph's structure, enabling richer relational modeling in which the media itself can be the subject or object of a triple.
Early systems such as IMGpedia~\cite{DBLP:conf/semweb/FerradaBH17} pioneered this approach by introducing a large-scale linked dataset that incorporated visual similarity relations between images.
Meanwhile, Richpedia~\cite{DBLP:journals/bdr/WangWQZ20} expanded this paradigm by providing a more comprehensive graph of textual and image entities.
However, these contributions primarily represent static datasets or ontologies rather than specialized storage systems.
By relying on general-purpose triplestores, they treat multimedia content as external URIs, resulting in \quot{opaque entities.}

Hence, a fundamental limitation of these systems is their content-agnostic nature, which isolates media content from the query engine, preventing it from natively interpreting the media's internal structure and requiring external processing for feature extraction or segmentation.
Recent advancements have introduced solutions like multimodal knowledge hypergraphs~\cite{DBLP:conf/aaai/ZengJBL23} to model complex many-to-many relations across modalities.
Additionally, frameworks such as MR-MKG~\cite{lee-etal-2024-multimodal} utilize relation graph attention networks to align image-text embeddings, facilitating advanced multimodal reasoning in large language models.
However, there remains a significant gap for a unified content-aware store that enables the query engine to directly access media content for dynamic operations.
Our work with \megras{} represents a critical advancement by providing a dedicated storage layer specifically designed for node-centric multimodal knowledge graphs.
Unlike previous systems that separate the graph structure from the media files, \megras{} integrates the intrinsic content of the media directly into the graph's node structure, enabling the query engine to interpret it not just as a link but as a queryable entity.

\subsection{Multimedia Information Retrieval}

The field of multimedia information retrieval has long been characterized by the \quot{semantic gap}~\cite{895972}, the disparity between low-level features and high-level human concepts.
To bridge this gap, the entire domain has undergone a fundamental shift from early statistical models to modern paradigms that leverage high-dimensional vector representations for cross-modal tasks.
Historically, the semantic gap has been the primary bottleneck to retrieval accuracy.
Early breakthroughs in cross-modal retrieval, such as semantic correlation matching~\cite{DBLP:journals/pami/PereiraCDRLLV14}, attempted to address the gap by mapping different modalities into a shared latent space where semantic similarity could be measured.
While these advancements enabled systems to retrieve images based on text or vice versa, they remained largely \quot{content-agnostic} in their execution;
The retrieval engines functioned as black boxes that mapped inputs to precomputed embeddings without understanding the underlying relational context of the media.
This lack of structural depth made complex, multi-hop reasoning over multimedia content nearly impossible within a single retrieval step.

To address these limitations, recent research has moved toward more expressive query languages and \quot{content-aware} architectures.
Standard graph query languages like SPARQL were not designed for high-dimensional similarity operators or temporal logic required for modern multimedia analysis.
Extensions such as SPARQL-MM~\cite{DBLP:conf/esws/KurzSSSK14} introduced primitives for spatial and temporal filtering, but often relying on static, pre-materialized metadata, necessitating a disjointed \quot{retrieval-then-analysis} pipeline.
\megras{} positions itself as a critical advancement by operationalizing a \quot{query-time-analysis} paradigm.
By integrating a high-performance database that efficiently handles vectors directly with a relational graph engine, \megras{} enables native similarity searches and the dynamic generation of media segments within the query execution plan.
Furthermore, by borrowing from SPARQL-MM and integrating concepts such as Allen's interval algebra~\cite{DBLP:journals/cacm/Allen83}, it allows us to reason over temporal relationships at the query level, enabling the search for complex spatiotemporal patterns that were previously inaccessible to content-agnostic systems.
The segments to be reasoned over do not need to exist ex-ante in the graph, but can be generated dynamically using the \ums{}~\cite{DBLP:journals/tgdk/WilliBR24}.
This transition from isolated retrieval tasks to a unified, scalable multimedia analysis establishes \megras{} as the first engine to bridge the gap between structured knowledge and intrinsic multimedia content at scale.

\section{The MediaGraph Data Model}
\label{sec:MG}

This work builds upon decades of effort to structure and connect data.
The term semantic web was coined in 2001~\cite{hendler2001semantic}, envisioning a future where data is machine-readable and linked on a global scale, primarily through standardized technologies like RDF (Resource Description Framework).\footnote{\url{https://www.w3.org/RDF}}
More recently, Google popularized the notion of knowledge graphs~\cite{singhal2012introducing}, demonstrating the immense practical value of structuring real-world entities and their relationships for large-scale applications.

For the formalization, we follow the definition of a knowledge graph as described in \cite{DBLP:journals/semweb/FarberBMR18}, based on the RDF specification: A graph $\mathcal{G}_{\text{RDF}}$ is a finite set of triples.
First, we define the set of all possible RDF terms $\mathcal{T}$ as the union of three disjoint sets: the set of URIs $\mathcal{U}$, the set of blank nodes $\mathcal{B}$, and the set of literals $\mathcal{L}$.
An RDF triple $t$ is an ordered set $(s, p, o)$, where $s \in \mathcal{U} \cup \mathcal{B}$ is the subject, $p \in \mathcal{U}$ is the predicate, and $o \in \mathcal{U} \cup \mathcal{B} \cup \mathcal{L}$ is the object. 
Therefore, the set of all possible RDF triples is formalized as:
$$
\mathcal{G}_{\text{RDF}} \subseteq (\mathcal{U} \cup \mathcal{B}) \times \mathcal{U} \times (\mathcal{U} \cup \mathcal{B} \cup \mathcal{L})
$$

\begin{figure}[t]
    \centering
    \includegraphics[width=0.8\columnwidth]{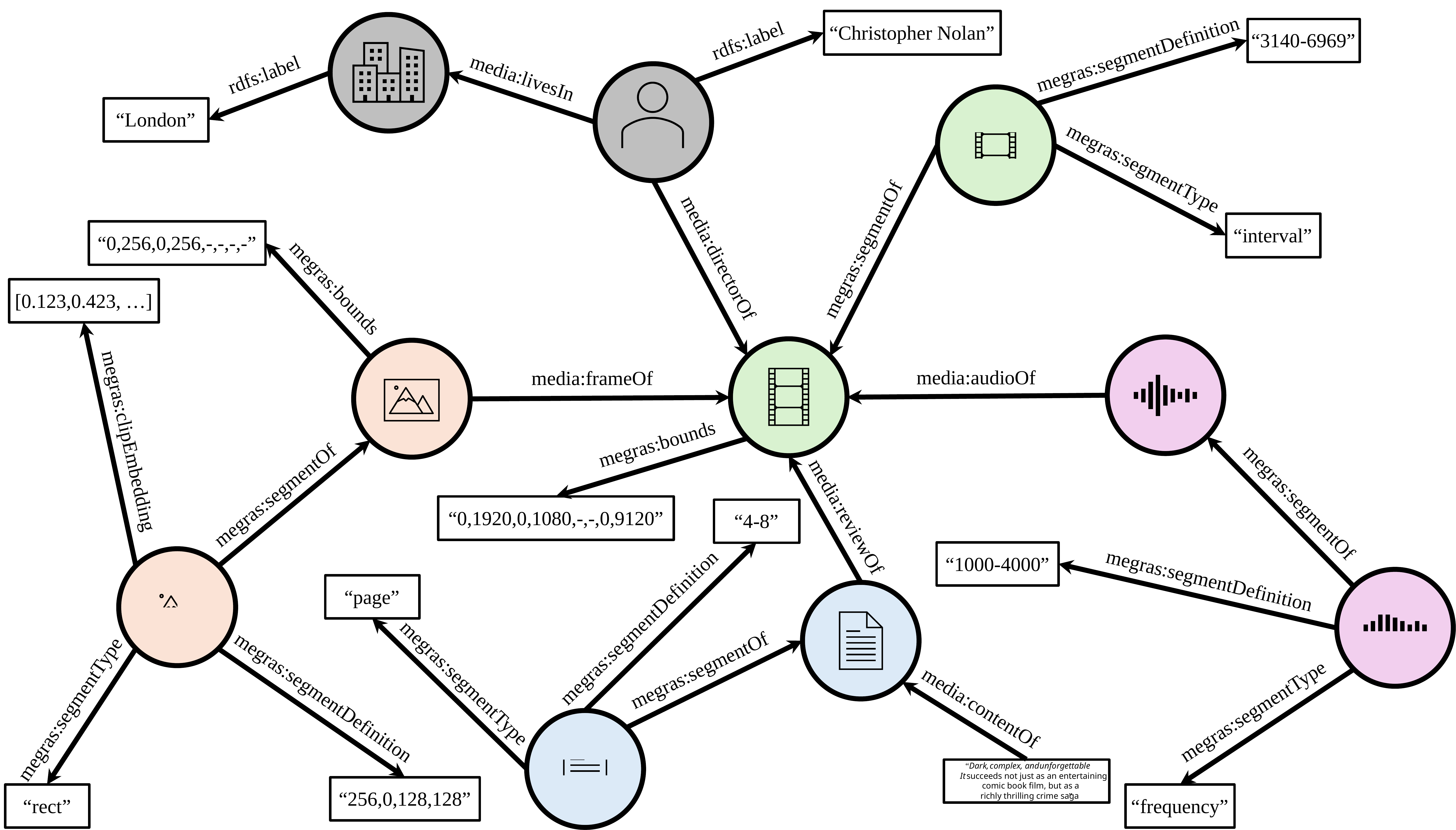}
    \caption{An example of a multimodal knowledge graph with the \mg{} Data Model. For simplicity, not all metadata literals are shown. The colors of the nodes represent their different types: orange for images, green for videos, violet for audio, blue for documents, and gray for generic.}
    \label{fig:mmkg}
\end{figure}

In accordance with this definition, we introduce the \mg{} Data Model to represent heterogeneous semantic data in a knowledge graph.
While traditional multimodal knowledge graphs store the multimedia documents either as opaque binary large objects (BLOBs) within a literal or referenced as external URIs, the \mg{} Data Model is designed to dissolve the barrier between structured and unstructured multimedia content.
Rather than treating it as atomic, unsearchable units, it moves to a \quot{content-aware} perspective, giving the query engine access, leveraging the intrinsic semantic richness.
By including the notion of Multimedia Document Nodes ($\mathcal{M}$), we elevate them to \quot{first-class} citizens in the knowledge graph

This paradigm shift means that every multimedia file is represented as an intrinsic node within the graph's URI-adressable space.
Therefore, the model ensures that the documents themselves become active participants in the graph's topology, capable of being the subject or object of any triple, while maintaining access to their content.
This approach enables a seamless integration of symbolic knowledge (concepts, entities) and perceptual data (images, video, audio) within a single, unified abstraction.
Unlike traditional systems that rely on a \quot{surrogate} metadata record to represent a file, we treat the file's URI as the primary node, removing layers of indirection and allowing for direct interaction between the query engine and the multimedia content.
This conceptual architecture enables a more holistic representation of multimodal data.

Hence, a \mg{} is defined as follows:
$$
\mathcal{G_{\text{MG}}} \subseteq (\mathcal{U} \cup \mathcal{B} \cup \mathcal{M}) \times \mathcal{U} \times (\mathcal{U} \cup \mathcal{B} \cup \mathcal{L} \cup \mathcal{M})
$$

An example of a multimodal knowledge graph with the \mg{} Data Model can be seen in \cref{fig:mmkg}: a video is at the center (represented with a green node) with annotations (gray nodes), such as its director, and file metadata (not entirely shown for simplicity).
It illustrates how multiple modalities can be utilized to capture different aspects of knowledge, using both actual \texttt{megras} and a fictional \texttt{media} ontology.
The latter is employed to link the video to related media: the associated audio (violet), a frame of the video as an image (orange), and a document (blue) containing a written review.
Unifying all these multimedia types in a single knowledge graph that also has access to their document contents yields advanced querying, analysis, and retrieval capabilities.


\section{MeGraS: the MediaGraph Store}
\label{sec:SY}

\begin{figure}[t]
    \centering
    \includegraphics[width=0.66\columnwidth]{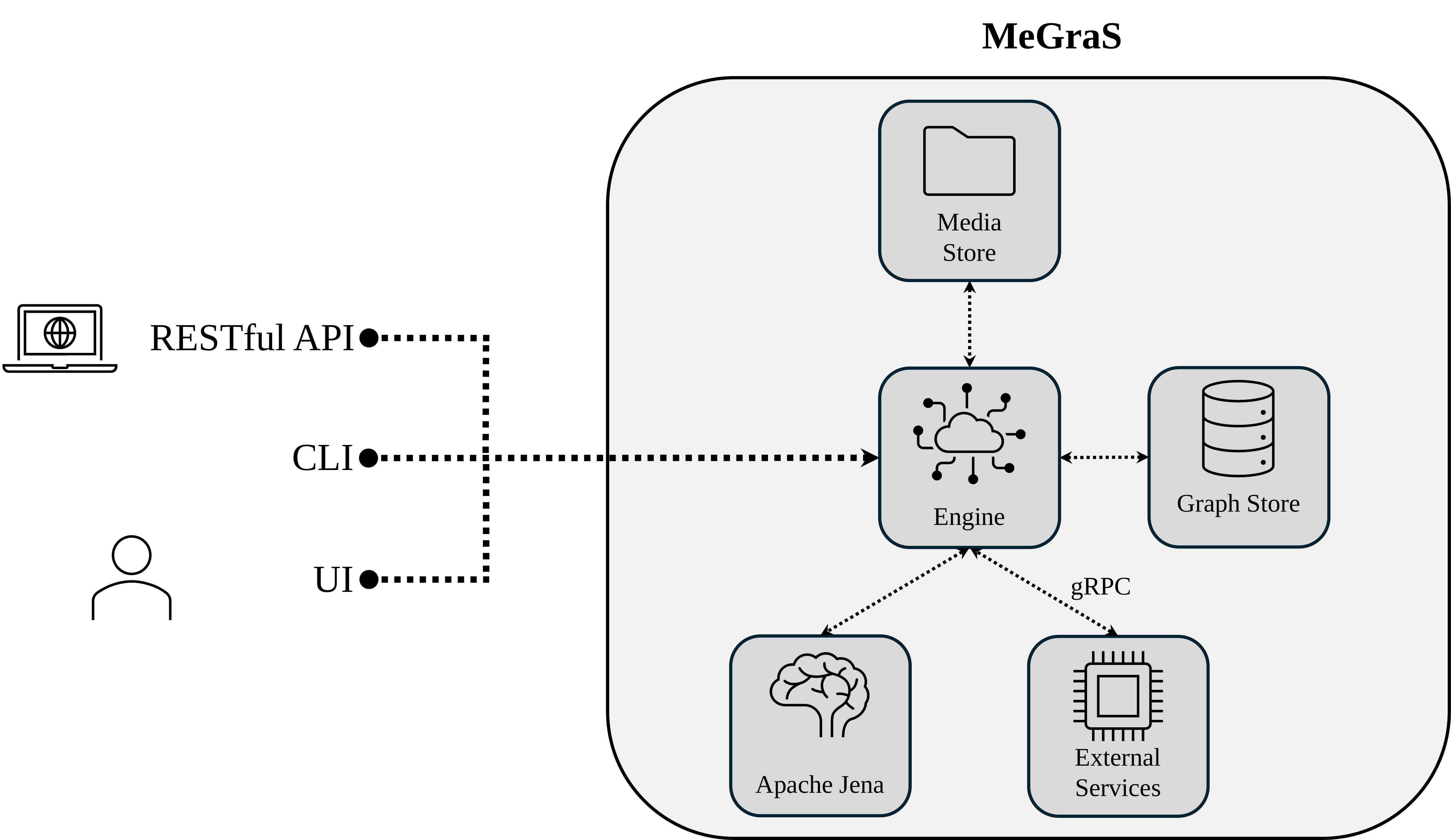}
    \caption{High-level system architecture diagram of \megras.}
    \label{fig:architecture}
\end{figure}

In this section, we explore \megras, our novel \mg{} Store, in detail.
First, we describe the core features of \megras: the underlying storage and query engine, the integrated multimedia segmentation, and the advanced querying capabilities.
We then present details of the current \megras{} implementation.
Finally, we discuss ways to interact with and manipulate the multimodal knowledge graph.

\cref{fig:architecture} provides a high-level overview of \megras's architecture.
It visualizes the various ways to interact with the system on the left, relating to the engine for querying and manipulating the graph at its heart.
It also shows the integration of Apache Jena\footnote{https://jena.apache.org/} for handling the SPARQL query language, as well as external services connected via gRPC.
Furthermore, it emphasizes the direct access to the media store that the query engine provides, and, finally, its interaction with the graph persistence store.

\subsection{Core Features}

At its heart, \megras{} is a query engine and a store for multimodal data, designed to bridge the gap between traditional knowledge representation and the richness of multimedia content.
While classic knowledge graph systems treat images, videos, or audio as opaque, external entities, \megras{} fundamentally integrates these media objects and their intrinsic features directly as graph nodes.
This innovative approach enables unparalleled query expressiveness, allowing for operations such as feature-based similarity search and dynamic segmentation to be performed natively within the graph framework.
We call this a paradigm shift from \quot{content-agnostic} to \quot{content-aware}~\anon[(Anonymous Reference)]{\cite{ruosch2026applications}}.

\subsubsection{Storage}

The storage architecture of \megras{} is engineered as a dual-component system, separating the management of semantic triples from the repository of raw multimedia content, but still unifying them in a single knowledge graph.
This decoupling is essential for maintaining high-performance query execution while handling the significant binary volume characteristic of multimodal data.
The architecture ensures that even though the \quot{logic} of the graph lives in a structured persistence layer and the \quot{substance} of the media is kept in the specialized store, both components are synchronized through a strict URI-based identification scheme.
The two components are shown in \cref{fig:architecture}: the media store above the engine and the graph store to the right.

\paragraph{Graph Store and Persistence Layer}
The persistence layer serves as the primary repository for \megras's structural data, consisting of triples.
To accommodate varying operational requirements, \megras{} utilizes a modular backend design that interfaces with multiple database engines through a unified abstraction layer.
For high-performance production environments, PostgreSQL\footnote{\url{https://www.postgresql.org}} is available and enhanced by the \texttt{pgvector}\footnote{\url{https://github.com/pgvector/pgvector}} extension, which facilitates the storage and indexing of vectors alongside traditional relational data.
It allows the use of advanced indexing structures to enable efficient similarity searches within the query execution pipeline.
Alternatively, for workloads that are heavily centered on vector-space operations, Cottontail DB~\cite{DBLP:conf/mm/GasserRHS20} is supported, a database specifically optimized for multimodal retrieval.
Finally, for smaller-scale applications, development, or demonstration purposes, \megras{} provides a lightweight, file-based backend that manages triples as tab-separated values, loading the entire graph into memory for low-latency pattern matching.

\paragraph{Media Store and URI Generation}
The media store functions as a dedicated repository for raw multimedia files, implementing a content-addressable storage paradigm that ensures data integrity and global uniqueness.
Unlike traditional systems that identify files by arbitrary names or paths, \megras{} derives a document's identity directly from its binary content using a rigorous hashing process.
Upon ingestion, every document is processed using the MultiHash\footnote{\url{https://github.com/multiformats/multihash}} standard, a self-describing hash format that prepends a header to the hash digest, specifying the hashing algorithm used (typically SHA3-256), ensuring the system remains future-proof and algorithm-agnostic.
This hashing process guarantees that identical files will always produce the same identifier, regardless of their source or original filename, facilitating automatic deduplication and preventing the proliferation of redundant nodes in the graph.

The file identifier is then derived from the computed hash by converting it to a URI-safe base64 representation, which is also trimmed (to remove the padding character \quot{=}).
The original multihash consists of $34$ bytes ($32$ from the SHA3-256, $2$ for the header).
This results in a string of $46$ characters ($\lceil (34 \text{ bytes} \times 8 \text{ bits per byte}) / 6 \text{ bits per character} \rceil = \lceil 45.33 \dots \rceil = 46 \text{ characters}$) that is used as the URI, serving as the permanent, resolvable address for that media node across the entire system.
Because these URIs are derived from the content itself, they provide a stable link between the semantic graph and the physical file.
When the query engine or an external client requests a media node, the media store resolves this URI to its corresponding binary stream.
During this process, the store can also perform canonical normalization, such as converting various image formats into a standard representation to ensure consistent feature extraction.
By utilizing this content-derived identification scheme, \megras{} ensures that every media node in the graph is not just a conceptual placeholder but a direct, verifiable reference to a specific piece of information.

The URI is also employed to build the raw media store, with the path derived from it.
The pairs of the first four characters are used to determine the nested subdirectories (e.g., \texttt{(a1/b2/)}, in which the raw bytes are written as a file named after the full hash (e.g., \texttt{basePath/a1/b2/<full-hash>}).
Additionally, a descriptor file is saved to the same location, storing essential metadata in JSON format.

During ingestion, \megras{} also writes a set of default metadata relations to the triple store: bounds, canonical ID, canonical MIME type, file name, media type, raw ID, and raw MIME type.
Additionally, more detailed image extraction is available via the Exchangeable Image File Format (Exif).
All this information is then available in the graph.

\subsubsection{Query Engine}

The \megras{} query engine is engineered around the \texttt{QuadSet} abstraction, a functional interface that extends the traditional RDF triple model.
It does so by incorporating a fourth component: a unique statement identifier.
\megras{} utilizes this feature to provide a stable, internal reference for every statement in the graph.
This design choice enables high-performance indexing, efficient statement-level caching, and robust deduplication via hashing.
By treating every statement as a distinct, addressable quad, the system can perform direct lookups and maintain a high degree of structural integrity during complex retrieval operations, allowing specific statements to be referenced, indexed, or updated with precision.

This abstraction layer does not merely serve as a storage interface; it defines the core retrieval capabilities of \megras{} by treating multimedia primitives as first-class operations.
The \texttt{QuadSet} provides a unified API that encompasses both semantic pattern matching and specialized multimedia retrieval, such as nearest-neighbor searches over high-dimensional vector embeddings and full-text filtering.
By standardizing these operations at the storage level, \megras{} can interface with the heterogeneous backends described above, while presenting a consistent, immutable graph view to the higher-level query processor.

At its most fundamental level, querying in this architecture relies on a triple-matching paradigm, where Basic Graph Patterns (BGP) are resolved by matching subjects, predicates, and objects against the stored quads.
During execution, the engine evaluates these patterns to bind variables to concrete values found within the graph.
This process typically proceeds through a series of sequential lookups: once a variable is bound in a preceding triple pattern, the value is used as a constraint for the subsequent \quot{hop} in the graph traversal.
This iterative matching continues until all patterns in the BGP are satisfied, effectively navigating the network of relationships to reconstruct the requested subgraph.

To facilitate standard-compliant access and enable intuitive and expressive querying, \megras{} implements a custom execution pipeline that extends the Apache Jena\footnote{\url{https://jena.apache.org/}} framework to support the semantic web query language SPARQL.
By replacing Jena's default sequential execution logic with a specialized batching engine, \megras{} adopts a bulk-processing paradigm for SPARQL queries.
In this model, the engine collects variable bindings from preceding query segments into a single context and executes them as a single, batched call to the underlying \texttt{QuadSet}.
These results are then joined in memory using optimized quad-indexing structures, ensuring that query latency remains manageable even as the complexity and dimensionality of the multimodal graph increase.

The efficiency of this execution model is further refined through a suite of algebraic optimizations that dynamically adapt to the query structure.
Before a BGP is executed, the engine reorders triple patterns based on selectivity heuristics, ensuring that the most restrictive patterns, such as those containing constant objects, are evaluated first to prune the search space.
Furthermore, the engine facilitates the pushdown of \texttt{LIMIT}, \texttt{PROJECTION}, and \texttt{JOIN} operators toward the storage layer.
By evaluating these constraints as early as possible, the engine can achieve early termination for limited result sets and minimize memory overhead by restricting variable binding to only those elements necessary for the final projection.
This combination of a statement-centric data model and an optimized batching executor provides the robust performance required for complex, multi-stage queries over heterogeneous multimodal data.

To further bridge the gap between abstract graph patterns and the physical properties of multimedia, \megras{} augments SPARQL with a comprehensive library of custom functions.
Currently, these extensions allow for the direct evaluation of geometric, temporal, and feature-based properties within the filter logic.
Specifically, the engine introduces spatial and temporal functions to calculate segment areas, bounding box centers, and the intrinsic dimensions of media segments across the 4D (XYZT) space.
By registering these functions under a dedicated namespace and providing them with direct access to the underlying storage abstractions, \megras{} empowers users to express complex, content-aware retrieval criteria that transcend the limits of traditional keyword or relationship-based searching.

\subsubsection{Unified Multimedia Segmentation}
\label{ssec:ums}

\begin{figure}[t]
    \centering
    \includegraphics[width=0.8\columnwidth]{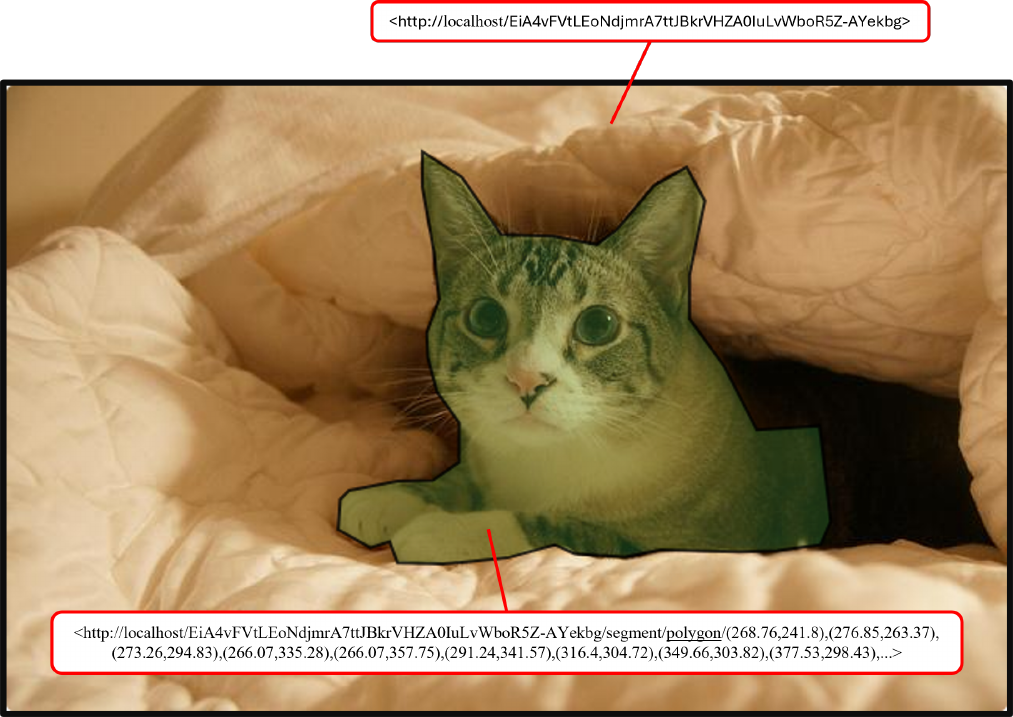}
    \caption{An example of a segmented image, taken from the COCO dataset~\cite{DBLP:conf/eccv/LinMBHPRDZ14}, with all parts being URI-addressable.}
    \label{fig:segmented}
\end{figure}

In another step to overcome the limitations of traditional \quot{content-agnostic} knowledge graphs, \megras{} incorporates the \ums{} (UMS)~\cite{DBLP:journals/tgdk/WilliBR24} model to break the previously atomic nature of multimedia documents.
It allows decomposing media files into addressable components and treats the segments as \quot{first-class citizens} that are natively processable within the graph.
The UMS model provides a formal mathematical foundation for defining segments across disparate media types by distinguishing their coordinate (the \quot{where} and \quot{when}) and description (the \quot{what}) dimensions.
An overview of all operations is shown in \cref{tab:ums}.
Hereby, UMS defines two main categories.
\textit{Filters} select a subset of the media's coordinate dimensions, such as specific spatial regions in an image or a temporal snippet of a video.
\textit{Reductions}, on the other hand, retain only specific portions of the document's description dimensions, such as extracting an audio track from a video or a specific color channel from an image.
To ensure these segments are accessible for querying, \megras{} utilizes an RDF-compatible URI naming scheme:
\begin{displayquote}
    <server address>/<canonical id>/\textbf{segment}/<segment type>/<segment description>
\end{displayquote}

An example of a segmented image is depicted in \cref{fig:segmented}.
For the segment, the URI shown includes its description, which consists of the type (\uline{\texttt{polygon}}) and a definition that is a list of points.
Furthermore, the example knowledge graph in \cref{fig:mmkg} illustrates the segments for all the diverse multimedia documents.
It shows an extracted interval of a video, selected pages of a document, and a rectangular segment of an image.
Finally, \cref{fig:about} shows how the segments are stored and visualized directly in \megras.

\begin{table}[h]
    \centering
    \caption{Overview of the operations based on the Unified Multimedia Segmentation model~\cite{DBLP:journals/tgdk/WilliBR24}.}
    \label{tab:ums}
    \begin{tabular}{@{} l l c l l @{}}
        \toprule
        \textbf{Category} & \textbf{Basis} & \textbf{Dimension} & \textbf{Type} & \textbf{Definition} \\ \midrule
        
        \multirow{16}{*}{\textbf{Filter}}
         & \multirow{12}{*}{Boundary}
                       & \multirow{5}{*}{1} & \texttt{TIME} & Temporal interval \\ \cmidrule(l){4-5} 
                 &     &                     & \texttt{CHARACTER} & Textual character-based interval \\ \cmidrule(l){4-5} 
                 &     &                     & \texttt{PAGE} & Document page-based interval \\ \cmidrule(l){4-5} 
                 &     &                     & \texttt{FREQUENCY} & Spectral range \\ \cmidrule(l){3-5} 
                 &     & \multirow{6.5}{*}{2} & \texttt{RECT} & Rectangular area \\ \cmidrule(l){4-5} 
                 &     &                     & \texttt{POLYGON} & Vertex list \\ \cmidrule(l){4-5} 
                 &     &                     & \texttt{PATH} & SVG-compliant path definition \\ \cmidrule(l){4-5} 
                 &     &                     & \texttt{BSPLINE} & Basis spline control points \\ \cmidrule(l){4-5} 
                 &     &                     & \texttt{BEZIER} & Bézier spline \\ \cmidrule(l){2-5} 
         & Cut Equation & n                  & \texttt{CUT} & Algebraic hypersurface \\ \cmidrule(l){2-5} 
         & Masking      & n                  & \texttt{MASK} & Binary values or base64 data-URL \\ \cmidrule(l){2-5} 
         & Space-Filling & n                 & \texttt{HILBERT} & Hilbert curve order and index ranges \\
        \midrule
        \multirow{5}{*}{\textbf{Reduction}} & \multirow{5}{*}{Dimensionality}
                       & \multirow{2.5}{*}{1} & \texttt{WIDTH} & Width interval \\ \cmidrule(l){4-5} 
                 &     &                     & \texttt{HEIGHT} & Height interval \\ \cmidrule(l){3-5}        
                 &     & \multirow{2.5}{*}{n} & \texttt{CHANNEL} & Logical media track/channel selection \\ \cmidrule(l){4-5} 
                 &     &                   & \texttt{COLOR} & Specific color channel selection \\
    \bottomrule  
    \end{tabular}
\end{table}

\subsubsection{Vector-based Analytics}

Beyond symbolic relationships, \megras{} incorporates high-dimensional feature vectors as native components of its quad-based architecture.
This integration allows these mathematical representations to be stored and queried as intrinsic graph objects, facilitating a direct connection between representations of structured semantic data and the encoded perceptual features of multimedia documents.
To support a wide range of analytical needs, \megras{} provides a tiered type system: \texttt{DoubleVectorValue} for high-precision scientific modeling, \texttt{FloatVectorValue} for standard machine learning applications, and \texttt{LongVectorValue} for discrete categorical or hashed features.
These types are designed for high-performance retrieval, mapping directly to optimized vector-space structures, natively for Cottontail DB and in PostgreSQL through \texttt{pgvector}.

The utility of these stored vectors is realized through a flexible similarity search framework centered on the \texttt{QuadSet} interface.
By implementing k-nearest neighbors (kNN) queries, the system can identify the most similar quads to a query vector or, through inversion, the most dissimilar outlier.
These searches utilize standard distance metrics such as Cosine similarity, Euclidean distance (L2), and inner product, and can be scoped to specific predicates.
This allows for nuanced operations, such as finding visually related images or identifying semantically aligned text snippets within a broader graph context.
By exposing these capabilities through the API and extended query-language support with custom SPARQL functions, \megras{} enables hybrid workflows that unify traditional graph traversals and content-based similarity analysis into a single, holistic retrieval process.

\subsubsection{Dynamic Knowledge Generation}

Beyond static data retrieval, \megras{} implements a framework for dynamic knowledge generation that automatically infers and computes relationships between entities without requiring them to be explicitly created, annotated, or materialized in the graph.
This approach ensures both storage efficiency and data consistency by leveraging two complementary mechanisms: \textit{implicit relations}, which exist between two graph nodes and are resolved at query time based on their inherent properties, and \textit{derived relations}, which are generated on-demand from raw (multimedia) data and result from predefined functions~\anon[(Anonymous Reference)]{\cite{Ruosch:2025:LSC}}.

\textit{Implicit relations} allow the system to recognize connections that exist purely by virtue of an object's characteristics, such as spatial position, temporal extent, or structural hierarchy.
Instead of storing every possible link, the system evaluates these relationships during query execution, enabling real-time discovery of spatial overlaps, structural parent-child links, and semantic similarities.
To facilitate these operations within standard workflows, \megras{} implements the relations described by SPARQL-MM~\cite{DBLP:conf/esws/KurzSSSK14}, allowing for standardized multimedia query patterns.
This allows retrieval based on topological relationships among spatial objects, including containment, overlap, direction, and boundaries.
Similarly, Allen's Interval Algebra~\cite{DBLP:journals/cacm/Allen83} can be expressed through \texttt{TemporalValue}s and comparison operators (<, >, ==, etc.) for temporal or sequential relationships.
Likewise, similarity relations are defined, such as finding near-duplicates or k-nearest neighbors.
However, none of these \textit{implicit relations} are persisted in the graph, as they may change based on criteria or node modifications (changes, additions, or deletions).
This ensures that the graph remains highly responsive to changes, as relationships are always inferred from the current state of the objects.

\textit{Derived relations} play their complementary role, transforming raw multimedia content into actionable graph data on the fly.
When a specific relation is requested that is not already available in the store, \megras{} triggers a transformation pipeline to extract high-level information-—such as conceptual embeddings, visual features, textual content, or document structures.
While these are generated on demand to bridge the gap between unstructured media and structured semantic networks, the resulting data is then persisted back into the graph.
This hybrid approach avoids the massive storage overhead of comprehensive preprocessing while ensuring that computation is incurred only once, thereby optimizing the performance of subsequent queries.
The result is a highly extensible environment where the graph's intelligence can be dynamically expanded and solidified as it is explored.

\subsubsection{External Services}

To provide access to sophisticated Machine Learning and Deep Learning methods for content analysis and feature extraction, \megras{} relies on a set of derived relations.
Unlike traditional static properties, these relations do not necessarily represent pre-stored data; instead, they act as dynamic bridges to external computational services.
When a user includes one of these specific relations in a query, the engine determines whether the object value needs to be computed in real time or retrieved from the graph store.
Similarly, a set of custom functions can be used to invoke external services through native SPARQL syntax.

This integration is designed to be seamless: from the end user's perspective, querying for an embedding or an OCR transcript is syntactically identical to querying for a hard-coded metadata field, such as a file name or creation date.
The underlying complexity is abstracted away by the handler registry, including gRPC communication with external Python-based model servers and the management of high-dimensional data.

The primary relations powered by this external execution layer currently include:
\begin{itemize}
    \item \texttt{megras:clipEmbedding}:
    This relation links a media node to its high-dimensional vector representation generated by the pretrained CLIP (Contrastive Language-Image Pre-training) model~\cite{DBLP:conf/icml/RadfordKHRGASAM21}.
    This feature can then be used by the query engine to perform cross-modal similarity searches (e.g., matching text to images) without the user needing to manage the underlying vector space or model inference.
    
    \item \texttt{megras:ocr}:
    This relation provides a direct link between a visual media node and its textual content.
    If the text has not been previously extracted, the query engine automatically invokes an external Optical Character Recognition service to derive the value.
    This allows for full-text search operations over images and scanned documents to be expressed naturally within a standard graph query.
    
    \item \texttt{megras:documentModel}:
    This relation is used to access the structural hierarchy of complex documents.
    It triggers an external analysis service to parse pages, sections, paragraphs, figures, and tables into a structured format.
    This enables the engine to navigate the internal \quot{skeleton} as if it were a collection of native graph nodes and link them to the original document.
\end{itemize}

By modeling these sophisticated feature extraction tasks as relations rather than external function calls, \megras{} ensures that the power of deep learning is natively accessible to the query engine.
This approach allows the system to maintain the expressive power of SPARQL while providing a \quot{content-aware} retrieval experience where the distinction between materialized and derived data is invisible to the user.

\subsection{Implementation}

This subsection covers the details of \megras's implementation, including its technology stack, availability, and installation.

\subsubsection{Technology Stack}

\megras{} is implemented as a modular, extensible open-source software framework that leverages modern, widely adopted technologies to enable easy integration.
The core system is written in \texttt{Kotlin} and runs on the \texttt{Java Virtual Machine}, providing the benefits of concise syntax while maintaining full \texttt{Java} interoperability.
The selection of the technology stack is strategically driven by the requirements for efficient multimodal graph persistence, advanced querying, and seamless integration with state-of-the-art machine learning models and other downstream applications.

\begin{itemize}
        \item \textbf{Build and Management}:
        For comprehensive dependency management and build automation, \megras{} relies on \texttt{Gradle} utilizing the \texttt{Apache Groovy Domain-Specific Languages}.
        This choice simplifies the integration of both JVM-native libraries and external system dependencies.
        
        \item \textbf{Data Persistence}:
        Multiple options are available for persisting knowledge graphs.
        For large-scale deployment, \texttt{PostgreSQL} is utilized for its stability and scalability in production environments in conjunction with the \texttt{pgvector} extension and the \texttt{JetBrains Exposed Object-Relational Mapping}.
        The integration of \texttt{pgvector} is crucial because it enables native, efficient vector operations. 
        \texttt{Cottontail DB}~\cite{DBLP:conf/mm/GasserRHS20} is also available as an alternative backend.
        The third option is the file-backend, whereby triples are written in RDF-style as tab-separated values.
        In this mode, the entire graph is loaded into memory and retained there, trading off memory limitations for faster querying.
        
        \item \textbf{External Services}:
        To incorporate computationally intensive tasks, \megras{} uses \texttt{gRPC} for establishing high-performance remote procedure calls.
        This provides a standardized interface for connecting the \texttt{Kotlin} core to external services.
        In the current iteration, we have implemented a \texttt{Python} server that runs in parallel to perform the necessary services.
        Examples include access to a dedicated optical character recognition library or \texttt{transformers}, which allows the system to leverage the rich ecosystem of pre-trained models.
        
        \item \textbf{Interfacing}:
        The main interaction layer is exposed via a \texttt{RESTful API} built with the lightweight \texttt{Javalin} web framework, prioritized for its ease of use and interoperability.
        For data ingestion and administration, the \texttt{Clikt} library provides comprehensive command-line interface support.
        
        \item \textbf{Semantic Processing}:
        \megras{} utilizes \texttt{Apache Jena} for robust, extensible \texttt{SPARQL} handling, which is then translated into the internal query processing engine via dedicated data representation wrappers.
        They are also used for retrieving the bindings for the \texttt{SPARQL} results.
        
        \item \textbf{Documentation}:
        \megras{} provides two manually curated documentation files: the README and a getting-started guide.
        The API documentation is automatically generated and maintained through the integration of the \texttt{OpenAPI Specification} and the \texttt{Swagger} framework, ensuring developers have up-to-date resources for system interaction.\footnote{\label{fn:oas}\anon{\url{https://github.com/MediaGraphOrg/MeGraS/blob/master/docs/oas.json}}}
        Furthermore, the URIs of prominent predicates resolve to their descriptions for self-annotation.
        The overview is available on the website.\footnote{\anon{\url{https://megras.org/schema}}}
    \end{itemize}

\subsubsection{Availability and Installation}

\megras{} is distributed as open-source software under the MIT License, ensuring its availability for both quick demonstrations and long-term academic development.
As a collaborative platform for ongoing development and community contributions, the source code is hosted on GitHub.\footref{fn:repo}
This approach is a deliberate choice to foster reproducibility and to encourage the adoption and extension of \megras{} by the broader research community.

Furthermore, \megras{} is available as a preconfigured Docker container including all dependencies for seamless deployment.\footref{fn:docker}
This is the most straightforward method for a full-stack setup, ensuring a consistent, reproducible environment across different operating systems.

For developers and researchers who require more granular control, detailed instructions for manual setup are also available.
These guides cover the configuration of individual components, offering flexibility for integration into existing infrastructures.
The emphasis on accessibility and comprehensive documentation underscores our commitment to making \megras{} a foundational tool for future research in multimodal knowledge graphs.

\subsection{Graph Interaction and Manipulation}

\begin{figure}[t]
    \centering
    \includegraphics[width=0.99\columnwidth]{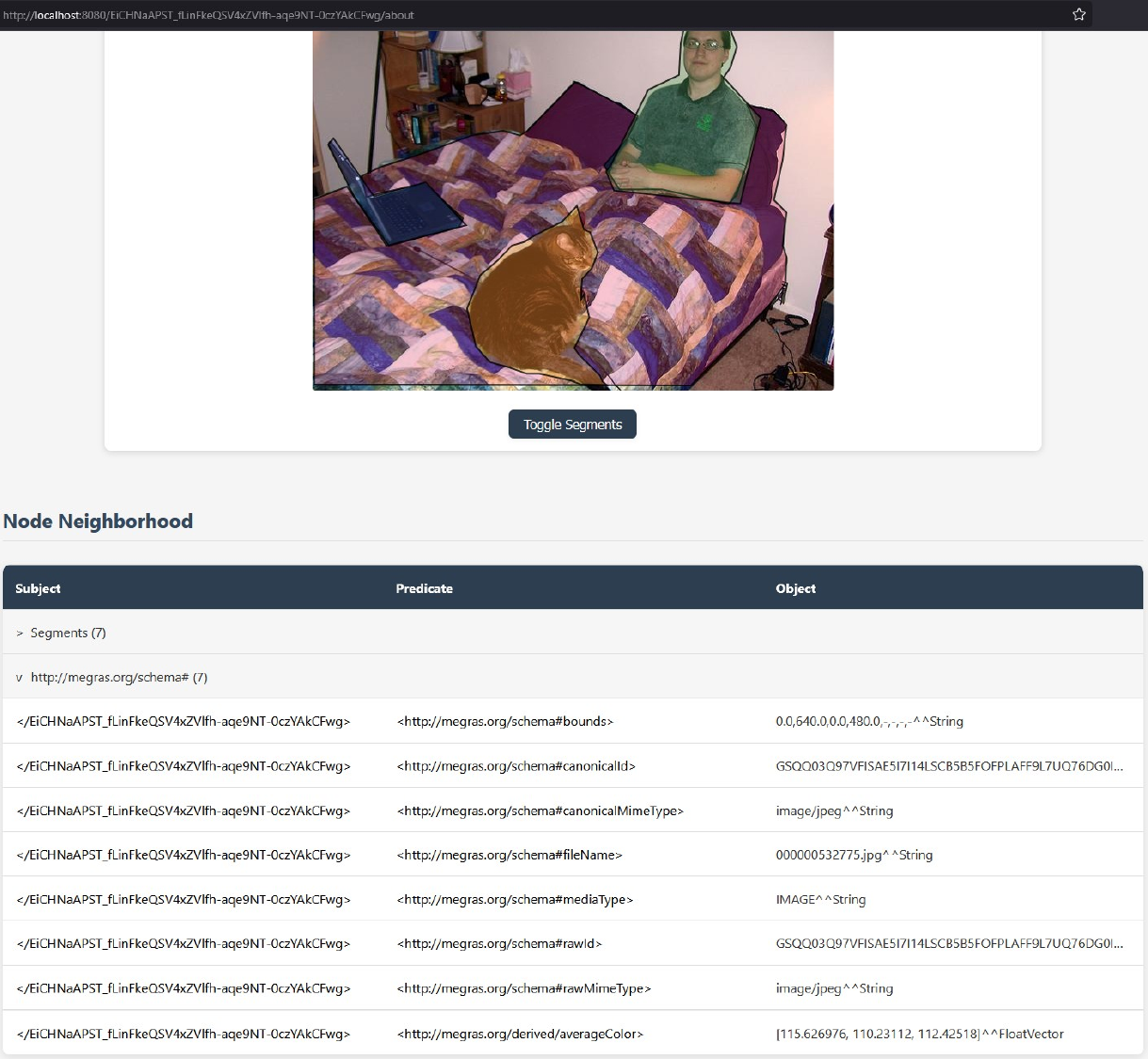}
    \caption{The data preview for an imported image, showing its URI and node neighborhood in the graph. The image and its segments are also rendered. The example is taken from the COCO dataset~\cite{DBLP:conf/eccv/LinMBHPRDZ14}.}
    \label{fig:about}
\end{figure}

As shown in \cref{fig:architecture}, there are three endpoints that allow for interaction with \megras{}. 
First, the RESTful Application Programming Interface (API) handles the bulk of graph operations, such as manipulation and querying, and is primarily designed for client applications.
Second, the Command-Line Interface (CLI) provides a straightforward way to perform simple data management tasks. 
Third, various user interfaces (UI) facilitate demonstrating \megras's features and can be used for showcases.

The methods of the RESTful API can generally be divided into three groups: data management, query, and \ums{} (UMS)~\cite{DBLP:journals/tgdk/WilliBR24}.
The detailed documentation can be found in the OpenAPI Specification.\footref{fn:oas}
The endpoints for data management provide fundamental methods for the lifecycle of both the knowledge graph and the associated multimedia objects: ingestion, access, deletion, and preview, including metadata.
The query API distinguishes three different types: pattern, content-aware, and SPARQL.
For basic path queries, the desired patterns can be given by individuals or sets of subjects, predicates, and objects as well as combinations thereof.
Content-aware queries operate in the feature vector space and can take the form of k-nearest neighbors or relevance feedback via iterative refinement, accepting both positive and negative examples.
Additionally, \megras{} provides a dedicated SPARQL endpoint implemented with Apache Jena.
It can be used to send pure SPARQL queries to our custom knowledge graph query engine, which will return the resulting bindings.
Finally, the API directly implements the principles of the UMS model, which is central to treating media components as first-class graph entities.
Every media segment is URI-addressable via a complex, generalized retrieval path that supports both segment fetching and algebraic composition.

The CLI mainly enables bulk data ingestion through two commands.
Using \texttt{add}, individual multimedia files or sets can be added to the knowledge graph.
Similarly, \texttt{import} allows loading a single TSV file with triples or sets thereof and adding them to the current knowledge graph.
Conversely, to export the graph from either database backend, \texttt{dump} creates a TSV file containing the RDF triples.
Finally, the \texttt{help} command can be used to obtain a list of all commands, their parameters, and descriptions.

The available UIs are a simple segmented utility, the option to add files or triples to the knowledge graph, and a data preview.
\cref{fig:about} shows an example of such an \quot{about page} of a knowledge graph node with its URI at the top.
The center part displays the multimedia document (in this case, an image) as served by our backend, including any segmentations with the colored, highlighted area.
The bottom half is dedicated to the node neighborhood in the knowledge graph, presented as a list of subjects, predicates, and triples.
The subject will always be the multimedia document for which the about page is opened, and the rows are grouped by predicate or its properties (e.g., prefixes).
The objects in the third column may be other graph nodes or literals.
Notably, the last row is a derived relation (\texttt{averageColor}) that has a vector as the object, as indicated by the \quot{FloatVector} suffix.
The ancestor neighborhood is not depicted in the figure; it is also displayed in a similar table.
Here, information about the nodes to which the current node is connected is shown, for example, facilitating the exploration of segment ancestors.

\section{Use Cases and Applications}
\label{sec:UC}

In this section, we discuss use cases and future applications of \megras.
First, we describe the \ligr{5} system, which used \megras{} as the underlying graph store and participated in the 2025 edition of the Lifelog Search Challenge.
Then, we explore how \megras{} can be used to model scientific papers, facilitating their exploration and information extraction.
Finally, we outline potential future applications.

\subsection{Semantic Search in a Multimodal Lifelog Dataset}

The practice of lifelogging~\cite{DBLP:journals/ftir/GurrinSD14} has led to the creation of vast and inherently multimodal datasets by continuously capturing individuals' daily experiences.
These collections, compiled from sources such as wearable cameras, biometric sensors, location logs, and the associated metadata, hold significant promise for augmenting human memory and enabling data-driven insights.
However, the sheer volume, diversity, and lack of structure pose a major obstacle to the efficient retrieval of lifelog data.

The annual \textit{Lifelog Search Challenge} (\textit{LSC})~\cite{lsc25}, the largest of its kind, aims to evaluate the performance of interactive retrieval systems in real time.
Each year, a cohort of teams participates to assess their proposed approach and its implementation.
The \textit{LSC} includes three task types that ar\textit{}e evaluated on the dedicated multimodal dataset.
These categories are \textit{known-item search}, \textit{ad-hoc search}, and \textit{question answering}.
In the \textit{known-item search}, a singular image is described more and more precisely over the course of five minutes, with a new hint becoming available every $30$ seconds.
The goal is then to locate that specific item.
For \textit{ad-hoc searches}, a single criterion is provided, and the participants have three minutes to find as many fitting images as possible.
Finally, \textit{question answering} requires submitting a textual reply that can be inferred from the information contained in the images.

The \textit{LSC} dataset comprises $725,000$ images from a first-person perspective, captured at regular intervals with a wearable camera over $18$ months.
They are organized in a hierarchical structure, grouped by day, with each item named by its timestamp.
For metadata, the logs of physical activities, biometrics (e.g., heart rate and step count), and GPS coordinates are provided.
Furthermore, two visual concepts are annotated: objects (detected using COCO~\cite{DBLP:conf/eccv/LinMBHPRDZ14}) and scenes~\cite{NIPS2014_19ea3982}.
The supplied additional information also includes semantic locations~\cite{10.1007/978-3-031-27818-1_54}.
Navigating this vast archive of data is very challenging, especially when considering the interactive aspect and, thus, the human in the loop.

\ligr{1}~\anon[(Anonymous Reference)]{\cite{Rossetto2020LifeGraph1}} was the first to propose a graph-based querying approach to the \textit{LSC}.
By processing the available data, in the form of images and accompanying annotations, into a knowledge graph, we captured and structured the information from the lifelogging data.
Furthermore, we then linked the graph nodes to external static knowledge bases and taxonomies to enhance the knowledge captured.
This enabled helpful operations, such as semantic expansion of concepts (e.g., finding not only cats but also all pets) and semantic similarity search (e.g., finding not only ships but also boats).
While the graph-based approach was novel, it did not perform as well as more established methods in the challenge.
Nevertheless, \ligra{} underwent multiple iterations, improving its performance and functionality, with a focus on both the technical architecture and the techniques used to enrich the data.

For the most recent \textit{LSC'25}, we developed \ligr{5}~\anon[(Anonymous Reference)]{\cite{Ruosch:2025:LSC}}, which comes with a specialized user interface to aid in and simplify the creation of queries to interact with a multimodal knowledge graph constructed from the dataset, stored in \megras{} as the backend.
Interactions happen solely through the SPARQL endpoint. 
Hence, we needed to provide an easy, intuitive way to construct these triple-pattern-matching statements, thereby facilitating image retrieval.
Based on a dedicated ontology, \ligr{5} retains the capture of the semantic information from previous \ligra s, while combining it with the power that \megras{} supplies.

Through \ligr{5}, we used \megras{} for all three task types that \textit{LSC} has to offer: KIS, QA, and ADHOC.
Known-item searches, or KIS for short, give participants 5 minutes to locate a specific item based on its description, which becomes more specific as the task progresses.
QA (question answering) also has a runtime of five minutes.
In it, participants use the dataset's information (images and metadata) to find the answer to a given question.
Finally, ADHOC tasks last only three minutes, and the assignment is to find as many images as possible that fit a given theme or description.
With \ligr{5}, \megras{} has demonstrated that it is well-suited to tackle challenges of semantic search in a multimodal lifelog dataset.

\begin{figure}[t]
    \centering
    \includegraphics[width=0.99\columnwidth]{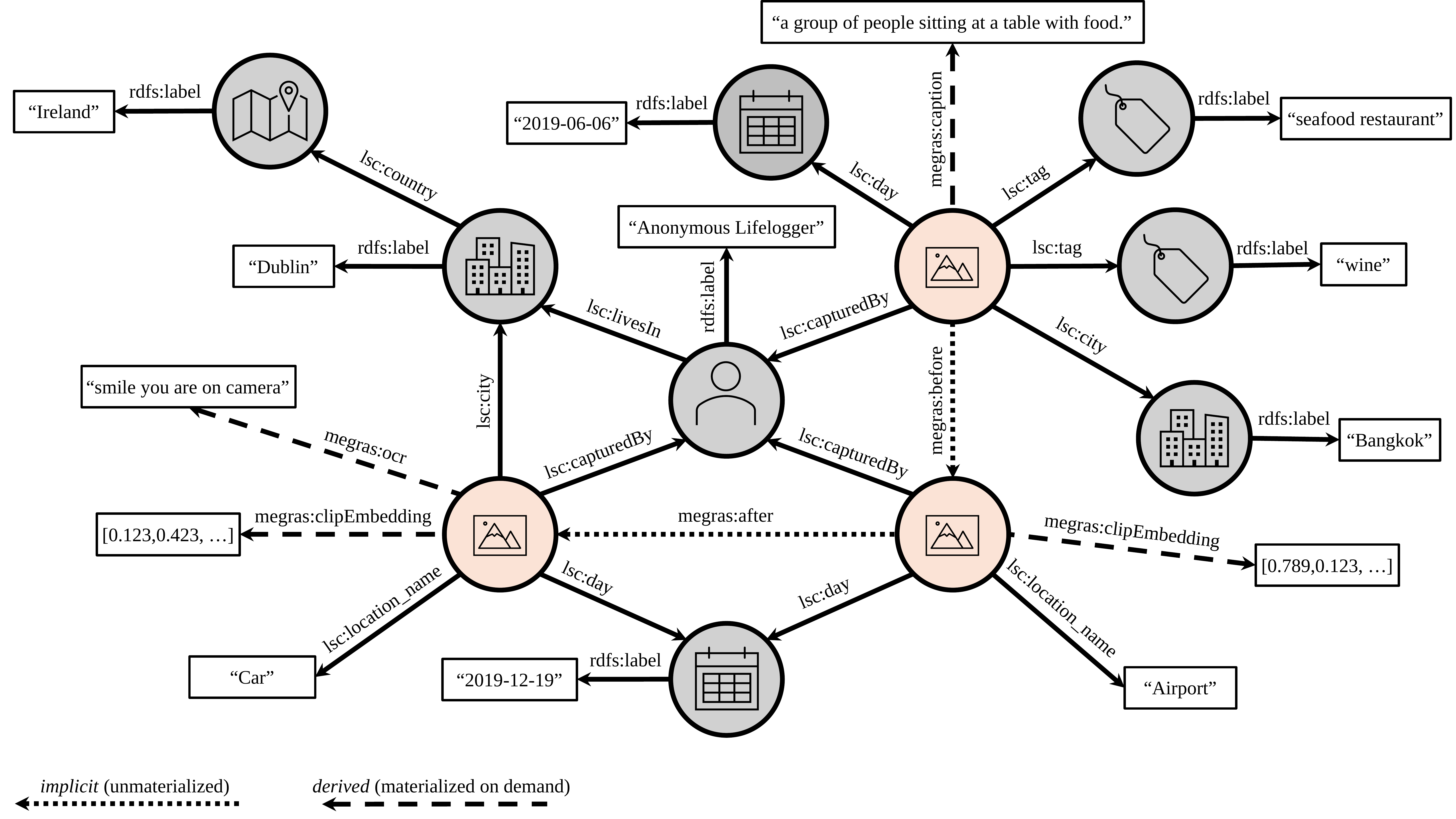}
    \caption{A minimal subset of the multimodal knowledge graph constructed from the Lifelog Search Challenge dataset~\cite{lsc25}.}
    \label{fig:lsckg}
\end{figure}

\paragraph{Case Study}
In order to illustrate how \ligr{5} works and leverages \megras, we present a case study based on an actual example.
\cref{fig:lsckg} shows a minimal subset of the multimodal knowledge graph constructed from the \textit{LSC} dataset.
It demonstrates how multiple modalities can be linked to enable advanced, effective semantic search across lifelog data.
The creator of the dataset, the anonymous lifelogger, is at the center, surrounded by images represented by orange nodes, which are linked to annotations and additional information.

Consider this real example of a QA task from \textit{LSC'25}~\cite{lsc25}:\footnote{\url{https://github.com/lucaro/LSC-Archive/tree/main/2025}}
\begin{displayquote}
\textit{I remember being in a taxi with a yellow sign taped to the back of a car headrest that said something like ‘smile you are on camera’. Where did I go to?}
\end{displayquote}
To figure this out, we first have to retrieve the corresponding image to find the time and date it was taken.
Then, we can expand the results based on this timeframe to any images captured afterwards.
Hence, several modalities are involved in this query: the image content (related to being in a taxi and the text shown) and metadata such as the date.

Since we already have some manual annotations in the dataset, such as tagged concepts, we can use them as a starting point for constructing our query to search the multimodal knowledge graph.
The full SPARQL query is shown in \cref{lst:lsc}.
First, we SELECT the image itself, the day it was taken on, and its identifier.
Then, we construct the triple patterns to match that data.
Furthermore, we put additional criteria into the WHERE clause.
We already have a \texttt{location\_name} for \texttt{Car}, so we will add that.
Next, we leverage one of \megras's derived relations, which enable processing multimedia documents on the fly, if the requested information is not already materialized in the knowledge graph.
In this case, we use the \texttt{ocr} predicate, which performs Optical Character Recognition and persists the extracted content if necessary.
We then FILTER on this result by looking for instances that contain the word \quot{smile.}

\begin{lstlisting}[language=SPARQL, caption={Example SPARQL query used in answering a question in the LSC'25~\cite{lsc25}.}, label=lst:lsc]
PREFIX lsc: <http://lsc.dcu.ie/schema#>
PREFIX megras: <http://megras.org/derived/>
PREFIX tag: <http://lsc.dcu.ie/tag#>

SELECT DISTINCT ?img ?day ?id
WHERE {
  ?img lsc:id ?id ;
    lsc:day ?day ;
    lsc:location_name "Car" ;
    megras:ocr ?ocr .
  FILTER(CONTAINS(LCASE(STR(?ocr)), LCASE("smile")))
}
ORDER BY ?id
\end{lstlisting}

Looking at \cref{fig:lsckg}, we see that the image node on the bottom left matches all these criteria.
Hence, it will also be among the retrieved results.
By triggering an expansion on the graph, we can retrieve the images taken shortly afterwards.
\megras{} also enables the use of its implicit relations, such as \texttt{after}, to place nodes in a temporal context.
This allows us to limit the result set to images taken on the same day, but with a timestamp after the root of our exploration, further simplifying the manual probing.
Browsing through these images eventually leads us to the bottom-right node, the first not tagged with \texttt{Car} anymore, giving us the answer: the lifelogger was on their way to the airport.

\subsection{Modeling And Exploring Scientific Literature}

Scientific knowledge is predominantly shared through research papers, which are typically distributed as immutable Portable Document Format (PDF) files.
These documents are notoriously difficult for machines to read and understand in a meaningful way, as they comprise a complex interplay of multimodal data, including text, figures, tables, and equations.
Conventional information retrieval systems often treat these papers as a flat corpus of text~\cite{gaizauskas-wilks-1998-information}, which fails to capture the underlying hierarchical structure and crucial relationships between their constituent parts.
This limitation makes it nearly impossible to perform complex, content-aware queries, which are essential for tasks such as systematic literature reviews~\cite{DBLP:conf/kes/SahlabKJW22}, automated knowledge base construction~\cite{DBLP:conf/doceng/Al-ZaidyG17}, or identifying novel research trends~\cite{DBLP:conf/esws/ZlochDDCZKMSOAD25}.

To capture this richness, we map the document structure into the \mg{} Data Model by treating each component as a first-class citizen in the graph.
Each paper is represented as a top-level document node, serving as the root for all intra-document content.
Using the \ums{} framework, individual contents, such as paragraphs, figures, and tables, are modeled as distinct segments of this root node.
This ensures a consistent representation in which a figure is not merely an image file but a node with spatial coordinates and semantic relations.
We model these through explicit intra-document relations (e.g., a paragraph node refers to a figure node) and inter-document relations, such as traditional citations or non-explicit links based on visual similarity between charts across different publications.

\begin{figure}[t]
    \centering
    \includegraphics[width=0.75\columnwidth]{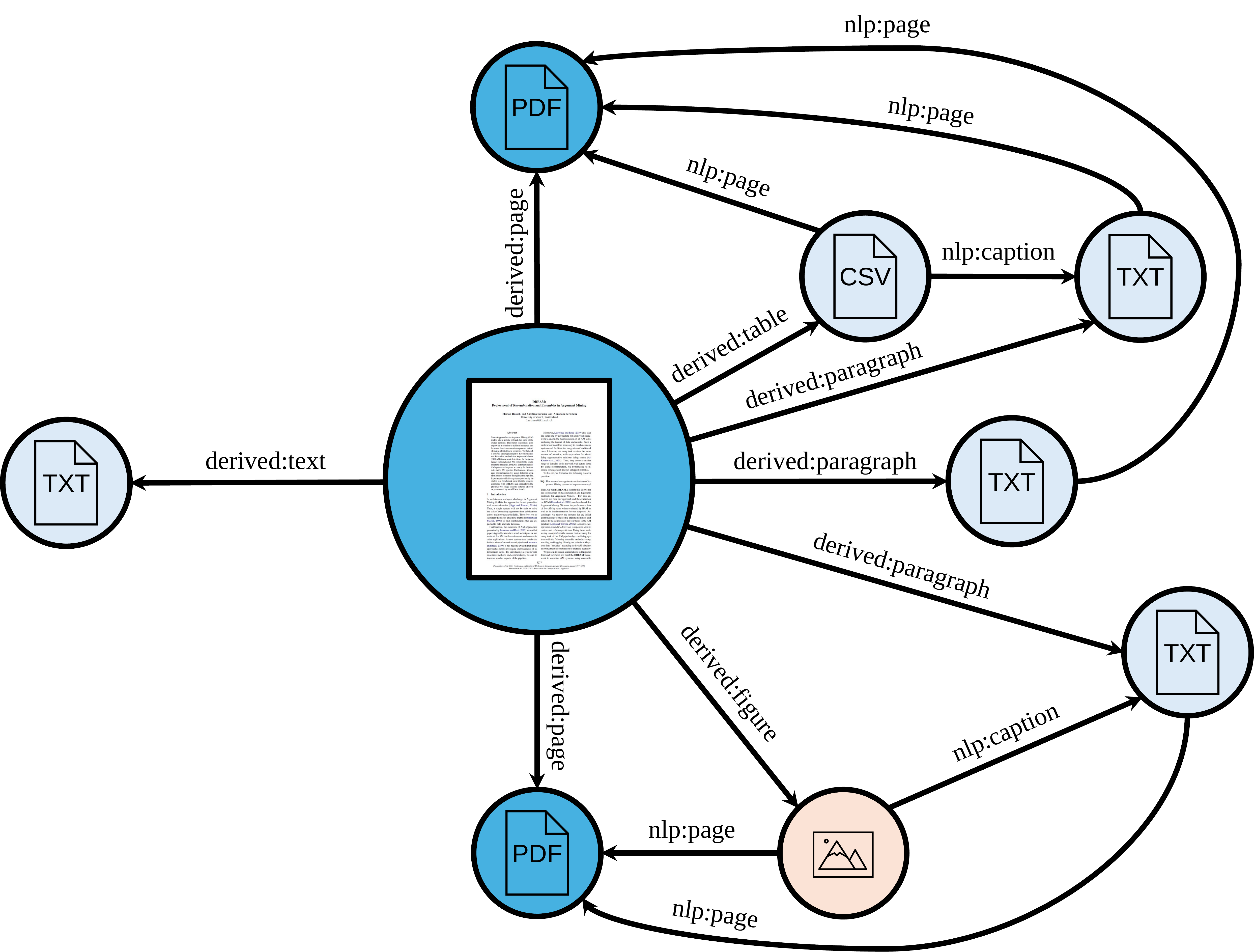}
    \caption{A minimal subset of the multimodal knowledge graph automatically constructed from a PDF file of a scientific paper.}
    \label{fig:nlpkg}
\end{figure}

\cref{fig:nlpkg} shows a minimal working example of a research paper dissected into several multimodal components.
This graph-based decomposition allows the system to move beyond simple keyword matching.
By populating the graph through automated extraction and identifying document components without manual annotation, we can validate the structural integrity of the representation and perform complex queries that respect the logical flow of the original paper.

\paragraph{Case Study}
To demonstrate the utility of this approach, consider a researcher performing a meta-analysis who needs to extract specific textual context surrounding a data visualization.
In a traditional system, searching for \quot{Figure 1} would return every instance of that string across a thousand-page corpus.
In \megras{}, we can leverage the structural modeling to be more precise.

Listing~\ref{lst:nlp} illustrates a query designed to retrieve the actual content and order of paragraphs that specifically reference a result table.
By querying the \texttt{derived:paragraph} relationship, the system ignores instances of "Figure 1" that might appear in the bibliography or header.
It filters for segments labeled "text" and retrieves their content along with their \texttt{nlp:ordinal} value.
This ensures that the extracted context maintains the narrative sequence intended by the authors, facilitating automated knowledge base construction or the synthesis of methodology sections across multiple papers.

\begin{lstlisting}[language=SPARQL, caption={Example SPARQL query for scientific literature analysis.}, label=lst:nlp]
PREFIX derived: <http://megras.org/derived/>
PREFIX nlp: <http://megras.org/nlp/>

SELECT ?p ?text
WHERE {
  <URI> derived:paragraph ?p .
  ?p nlp:ordinal ?ord ;
     nlp:label "text" ;
     derived:text ?text .
  FILTER CONTAINS(?text, "Figure 1")
}
ORDER BY ?ord
\end{lstlisting}




\subsection{Potential Applications}

The content-aware paradigm of \megras{} enables a broad range of applications that require deep integration of structured knowledge with raw multimedia data.
By elevating media and their segments to first-class graph entities, the system enables more sophisticated interaction across several domains.
Also, for an overview of more applications, we refer the inclined reader to \anon[(Anonymous Reference)]{\cite{ruosch2026applications}}.

\paragraph{Interactive Video Exploration and the Video Browser Showdown}
Similarly to its use for the LSC, the architecture of \megras{} can also show its practical utility in the context of interactive video retrieval challenges, such as the Video Browser Showdown (VBS)~\cite{DBLP:journals/mms/LokocABDGMMNPRSSSKSVV23}.
The VBS is an international competition that evaluates the efficiency of video search engines in solving KIS and ADHOC tasks under strict time constraints, akin to the LSC.
Previous work with \anon[Video Retrieval System]{VideoGraph}~\anon[(Anonymous Reference)]{\cite{DBLP:conf/mmm/RossettoBARPHB21}} has already established the feasibility of graph-based approaches, showing that information extracted from multiple video modalities can be combined with external knowledge bases to produce semantically enriched representations.
\megras{} builds on this by moving beyond the \quot{opaque blob} of videos, enabling native similarity searches and dynamically generating media segments during query execution.
This \quot{content-aware} approach provides the high retrieval precision needed for the rapid, iterative exploration required in competitive video retrieval settings.

\paragraph{Enhancing Search and Recommendation Systems}
\megras{} can significantly improve the relevance and accuracy of retrieval systems by moving beyond keyword-based indexing.
In a content-aware framework, a query for \quot{a sunny beach in Italy with good searching} is processed by linking the semantic concepts of \quot{beach,} \quot{sun,} and \quot{surfing} directly to high-dimensional visual and geographical features within the graph.
This allows the engine to return precise results, such as specific video segments rather than generic text articles.
This approach aligns with recent advancements in retrieval-augmented generation (RAG)~\cite{DBLP:conf/nips/LewisPPPKGKLYR020}, which demonstrate that combining pre-trained models with external document indices generates more factual and specific language than traditional baseline models.
In e-commerce, this enables recommendations based on a user's holistic interaction history, including viewed products, liked images, and video-watching patterns, all mapped within a unified relational space.

\paragraph{Improving Content Creation and Summarization}
The ability to link concepts across disparate modalities allows \megras{} to serve as a foundation for automated content generation.
By identifying and selecting relevant video clips, images, and text spans that correspond to specific semantic keys, the system can assist in generating automated summaries of complex documents, provided the appropriate external services are supplied.
In journalism, for instance, \megras{} can facilitate the creation of visually compelling news stories by automatically pairing long-form articles with the most relevant multimedia assets from a large-scale repository, ensuring that the generated content is both contextually and visually aligned.

\paragraph{Social Media Analysis}
In the context of social media, \megras{} facilitates the analysis of large, interconnected user-generated content to reconstruct and understand complex real-world events.
Rather than relying on simple text-based hashtag analysis, the system can correlate textual reports with visual evidence across a static archive of a specific event, such as a natural disaster or a public safety incident.
For example, it can link a specific GPS-tagged text post with visual features of a flood or fire extracted from user-uploaded videos, establishing a spatiotemporal relationship within the graph.
This allows researchers and authorities to perform expressive queries to identify \quot{visually confirmed} reports, establishing a more robust understanding of the situation than traditional content-agnostic analysis.

\paragraph{Urban Planning and Smart Cities}
The scalability of \megras{} makes it a powerful analytical tool for urban planners working with large-scale historical city data.
By ingesting and indexing a collection of data from city-wide sensors, such as street cameras, microphones, and public social media archives, it enables complex forensic queries.
An urban planner could analyze an archived month of city data to \quot{identify all road intersections where traffic noise exceeded a certain decibel level and was correlated with pedestrian congestion at specific times of the day,} combining audio and visual data.
This provides a data-driven foundation for policy decisions, such as zoning or infrastructure changes, by allowing planners to retrieve and visualize correlations across diverse multimedia modalities that were previously siloed.

\paragraph{Copyright and Intellectual Property}
Finally, \megras{} offers a robust framework for protecting intellectual property by creating a comprehensive knowledge graph of original content.
A query could be run to quickly find all instances where a specific logo from an image, a unique soundbite from an audio track, or a distinctive video sequence has been used without permission across various online platforms.
This provides a more robust, granular search than traditional content ID systems, leveraging \megras's ability to infer relations and perform similarity searches directly on intrinsic media features.

\section{Future Work and Limitations}
\label{sec:FW}

While we introduced the content-aware paradigm with \megras{} that shifts how multimedia is handled in knowledge graphs, several technical and conceptual challenges remain to be addressed in future development.
A primary area for improvement is system performance, particularly regarding the throughput of complex queries.
The current performance bottlenecks are localized to the interaction between the SPARQL interface and the Apache Jena framework, given the nature of the default wrapper.
Our current batching engine serves as a proof-of-concept for managing this overhead, but it is only an intermediate step.
Future research will move toward a sophisticated, cost-based query optimizer designed to balance the computational expenses.
Additionally, we plan to investigate streaming ingestion pipelines to support live multimedia analysis, enabling the graph to update dynamically as new media arrives.

Another inherent limitation lies in the deterministic nature of the current representation, owing to a fundamental property of knowledge graphs.
Storing extracted features as static triples fails to capture the probabilistic confidence levels in machine learning models.
To address this, future iterations will explore integrating provenance metadata and model versioning to ensure that derived knowledge is transparent and auditable.
This shift toward a more nuanced data model will also facilitate the development of a \quot{content-aware} reasoning engine capable of performing complex, rule-based inference directly on the extracted media content.
Such an engine would enable the system to automatically generate new knowledge without manual intervention, for example, by inferring social relationships from the repeated co-occurrence of individuals across different temporal segments.
However, the system remains subject to the \quot{semantic gap}; future work must explore how the graph's structural context can be used to disambiguate fuzzy multimedia signals.

Beyond these specific technical optimizations, we aim to refine the system into a truly generic framework capable of supporting arbitrary media types and a broader array of external feature-extraction services.
Furthermore, to bridge the gap for non-technical users, we plan to integrate natural-language-to-SPARQL capabilities, enabling more intuitive exploration of the multimodal knowledge graph through descriptive queries rather than formal syntax.
Beyond the individual store, a major direction for future work is the implementation of federated multimodal querying.
This would allow \megras{} to support distributed instances across different institutions, enabling a collaborative ecosystem for knowledge construction.
In this vision, cross-repository meta-analyses would be possible, allowing for the search and retrieval of media segments and their semantic relations across a global, federated, multimodal knowledge network.
This evolution would move the field closer to a truly decentralized and content-aware multimedia intelligence framework.

\section{Conclusion}
\label{sec:CO}

In this article, we introduced the \textbf{\mg{} Data Model}, a formal framework for content-aware multimodal knowledge graphs that overcomes a fundamental conceptual limitation of existing systems: their content-agnostic treatment of multimedia as opaque, external entities.
By elevating multimedia documents and their segments to first-class, URI-addressable citizens in the graph, the \mg{} model enables a class of querying, analysis, and inference that has previously been inaccessible; one in which the query engine has direct, native access to intrinsic media content rather than merely a reference to it.

We presented \textbf{\megras}~\anon[(Anonymous Reference)]{\cite{Rossetto:2025:MGR}}, short for \textbf{Me}dia\textbf{Gra}ph \textbf{S}tore, as a concrete, open-source instantiation of the \mg{} Data Model.
\megras{} validates the content-aware paradigm~\anon[(Anonymous Reference)]{\cite{ruosch2026applications}} through a high-performance storage architecture that decouples semantic triples from raw binary content while maintaining strict URI-based synchronization between the two.
The integration of the \ums{} and vector-based analytics enables native operations, such as feature-based similarity search and dynamic spatiotemporal reasoning, directly within the query execution pipeline.
We further detailed how \megras{} extends SPARQL to realize these capabilities, and described a framework for dynamic knowledge generation in which implicit and derived relations minimize storage overhead by computing complex relationships on demand and persisting them only when necessary.

Validation through diverse use cases, including the Lifelog Search Challenge~\cite{lsc25} and scientific literature modeling, demonstrates the \mg{} model's ability to facilitate cross-modal analysis and expressive querying at scale.
By releasing \megras{} as an open-source framework, we provide the research community with a practical foundation for developing advanced multimedia retrieval applications and exploring the intersection of structured knowledge graphs and deep learning.

Future work will focus on expanding the library of derived relations to cover more specialized modalities and on investigating further query-time optimizations to improve performance on massive, distributed multimodal knowledge graphs.
More broadly, we envision the \mg{} model as a step toward federated, content-aware multimedia intelligence: a landscape in which structured knowledge and perceptual content are no longer separated, but unified.

\begin{acks}
    This work was partially funded by the Swiss National Science Foundation through Project \href{https://data.snf.ch/grants/grant/202125}{``MediaGraph''} (Grant Number 202125).
\end{acks}

\bibliographystyle{ACM-Reference-Format}
\bibliography{bibliography}

\end{document}